\documentclass[12pt]{iopart}
\usepackage{iopams}
\usepackage{epsfig}

\begin{document}
\title{Statistical Physics of Biological Evolution}
\author{Joachim Krug}
\address{Institut f\"ur Theoretische Physik, Universit\"at zu K\"oln}
\eads{\mailto{krug@thp.uni-koeln.de}}
\date{today}
\begin{abstract}
This is an extended abstract of lectures delivered at the 4th Warsaw
Summer School on Statistical Physics in Kazimierz Dolny, June 25-July
2, 2011.
\end{abstract}
% \maketitle

\noindent
The mathematical theory of evolution is concerned with the changes in the genetic composition of 
populations that occur under the influence of the evolutionary forces of selection, mutation and 
demographic noise. In its focus on noisy, collective behavior in large ensembles of (relatively) 
simple constituents it displays many conceptual similarities to statistical physics, which have 
given rise to a fruitful interaction between the two fields in recent years. The aim of the lectures
delivered at Kazimierz Dolny was to 
introduce the basic concepts of the theory and to describe some recent work, with particular 
emphasis on results that are relevant to evolution experiments with microbial populations.
These brief notes summarize the main issues that were presented and provide a fairly extensive
list of references. 

\section{Basic concepts and evolutionary regimes}
\label{Intro}

The basics of mathematical population genetics were developed in the 1930's by Fisher, Haldane and Wright
\cite{Fisher,Haldane1927,Haldane1931,Wright1931,Wright1932}. These three names are generally associated with the 
'modern synthesis' of evolutionary biology, which unified the discrete nature of Mendelian heredity with the 
Darwinian picture of adaptation by small changes 
accumulated over long periods of time. Their key insight was that evolution should be viewed as a stochastic
phenomenon, where discrete, random mutational changes in single individuals
give rise to a seemingly deterministic adaptive process on the population level.
In this perspective, evolutionary theory is the statistical mechanics of genes. 

The standard model of adaptation on the population level is the Wright-Fisher model, which describes the 
evolution of a population of fixed size $N$ in discrete, non-overlapping generations. Mutations occur randomly at rate $U$ per generation, 
and selection is incorporated as a bias in the choice of offspring. Mathematically, the Wright-Fisher model can be defined
as a branching process conditioned on a fixed population size
\cite{Park2010}. 

An important elementary process is the 
\textit{fixation} of a new mutation which initially arises in a single individual. The probability of fixation can be computed
exactly for the branching process \cite{Haldane1927} as well as for the Moran model \cite{Moran}, a continuous time
process where individuals replicate and die one at a time. The most commonly used expression for the fixation probability
was derived by Kimura \cite{Kimura} in a continuum approximation based on a Langevin equation for the mutant frequency. 
A new mutation is most likely to go extinct during the early stage of the fixation process, and mutations that 
survive this initial stochastic regime are called \textit{established} \cite{Desai2007,MaynardSmith1971}. 

Depending on the population parameters $N$, $U$ and the typical selection coefficient $s$ describing the 
fitness advantage of the mutant, different evolutionary regimes emerge \cite{Gillespie1984,Park2010}. Selection is strong if $Ns \gg 1$ and weak if
$Ns \ll 1$. Moreover, when the time to fixation $t_\mathrm{fix} \sim s^{-1} \ln N$ is short compared to the time 
$t_\mathrm{mut} \sim (sUN)^{-1}$ between subsequent establishment events, mutations fix independently, 
whereas for $t_\mathrm{fix} > t_\mathrm{mut}$ they interfere (see Lecture \ref{CI} for further discussion of this regime).  

\section{Sequence space and fitness landscapes}
\label{SeqSp}

The genetic information is encoded in linear sequences of symbols drawn from a finite alphabet. On the microscopic level
the symbols stand for nucleotides forming DNA or RNA molecules, or for amino acids forming proteins; on the coarse grained
level of classical population genetics, they stand for different variants (\textit{alleles}) of a gene. For many purposes
it is sufficient to consider binary sequences, where the symbols merely indicate the presence or absence of a mutation
at a given genetic locus. The space of binary sequences of length $L$ is the $L$-dimensional \textit{hypercube} endowed 
with the \textit{Hamming distance} as the natural metric; the Hamming distance between two sequences is simply the number of letters in which they differ. 

Assuming that the fitness of an individual is completely determined by its genotype, fitness can be viewed
as a function on sequence space. This idea was first introduced by Haldane \cite{Haldane1931} and Wright \cite{Wright1932},
who also pointed out that the existence of multiple peaks in the fitness landscape was a likely scenario that could
obstruct the evolutionary process. Later Maynard Smith envisioned evolutionary trajectories as pathways in the space 
of amino acid sequences that are constrained to move from one viable protein to another \cite{MaynardSmith1970}.  
Recent years have seen a surge of renewed interest in the concept, triggered primarily by the availability of empirical 
data where fitness (or some proxy thereof, such as antibiotic resistance) is measured for all $2^L$ combinations of 
$L$ mutations (typically $L=4-8$), see 
\cite{Carneiro2010,Chou2011,deVisser1997,deVisser2009,Lozovsky2009,Poelwijk2007,Weinreich2006}. 

\section{Evolutionary accessibility of fitness landscapes}
\label{Acc}

In population genetic terminology, the notion of \textit{epistasis} refers to interactions between
different mutations in their effect on fitness. Of particular importance is \textit{sign epistasis}, which 
implies that a given mutation may be beneficial (increasing fitness) or deleterious (decreasing fitness)
depending on the presence of mutations at other loci. Fitness landscapes without sign epistasis
are simple, in the sense that they possess a unique fitness maximum, and fitness increases monotonically
along any path approaching the maximum \cite{Weinreich2005}. In the presence of sign epistasis at 
least some of the paths become inaccessible, in the sense that they include steps of decreasing fitness, 
but the existence of multiple fitness maxima requires a specific, stronger form of \textit{reciprocal}
sign epistasis \cite{Poelwijk2011}. 

The empirical studies described above in Lecture \ref{SeqSp} show that sign epistasis is prevalent
in nature, and it is therefore important to devise fitness landscape models that allow to quantify
this feature. From the point of view of statistical physics, a natural approach is to consider
random ensembles of fitness landscapes with prescribed statistics. In the simplest case random fitness
values are assigned independently to the genotypes, resulting in the House of Cards (HoC) model
first introduced by Kingman \cite{Kingman1978} and Kauffman and Levin \cite{Kauffman1987} in 
the genetic context; in the statistical physics of spin glasses this is known as Derrida's
Random Energy Model (REM) \cite{Derrida1981}. 

It is easy to see that the probability for a given genotype
to be a local fitness maximum is simply $1/(L+1)$ in the HoC model, and it can be shown that 
the distribution of the number of fitness maxima is asymptotically normal \cite{Baldi1989,Macken1989}. 
A simple combinatorial argument can also be applied to the question of evolutionary accessibility, 
showing that the expected number of fitness-monotonic paths to the global fitness optimum is 
equal to 1 irrespective of $L$ and of the initial distance to the peak \cite{Franke2011}. 
However, the full distribution of the number of accessible paths can only be explored by numerical
simulations. It is found to display large sample-to-sample fluctuations, with the majority of 
realizations (approaching unity for large $L$) having no accessible path spanning the entire landscape. 

Real fitness landscapes are not likely to be entirely uncorrelated, and different models with
a tunable degree of fitness correlations have been proposed. A classic example is 
the LK-model introduced by Kauffman and Weinberger \cite{Kauffman1989}, in which each of $L$
loci interacts randomly with $K$ other loci. For $K=0$ the landscape is non-epistatic, while for
$K=L-1$ it becomes equivalent to the HoC model. The statistics of local maxima in the LK-model has
been adressed analytically by probabilists \cite{Durrett2003,Limic2004}, but the properties of 
accessible mutational pathways has only been studied by simulations so far \cite{Franke2011}. In marked contrast to the HoC model, one finds an increase of evolutionary accessibility with increasing 
$L$ (in the sense that the likelihood to find at least one spanning accessible path to the global
fitness maximum increases) when the number of interacting loci $K$ is taken to be proportional to (but smaller than) $L$.

A second example of a tunably rugged fitness landscape is the Rough Mt. Fuji (RMF) model orignally
introduced in the context of protein evolution \cite{Aita2000}. In this model random fitness values
(as in the HoC model) are superimposed on an overall fitness gradient of tunable strength
$\theta$; in spin glass language, the model is equivalent to the REM in an external field. 
The problem of evolutionary accessibility in the RMF is closely related to the theory of records
in sets of independent random variables with a linear drift \cite{Franke2010}, and by exploiting
this connection analytic results for the expected number of accessible paths can be derived. 
One finds an increase of accessibility with increasing $L$ for any $\theta > 0$, 
reflecting the fact that the factorial growth in the number of possible pathways 
overwhelms the exponential decrease in the probability of any given pathway to be accessible
\cite{Franke2011}.

The quantitative measures of evolutionary accessibility developed in the model studies can be applied
to empirical fitness landscapes, with the aim of testing the models and estimating epistasis parameters
like $K$ and $\theta$. For this purpose it is useful to decompose the landscape into subgraphs
spanned by subsets of the total set of $L$ mutations under consideration, and to study the behavior
of the accessibility measures as a function of subgraph size. Applying this approach to a fitness data
set containing combinations of 8 individually deleterious mutations in the filamentous fungus
\textit{Aspergillus niger} \cite{deVisser1997}, it was found that the data are well described by
an LK-model with $K/L \approx 1/2$, or by an RMF-model with an intermediate value of $\theta$
\cite{Franke2011}.

\section{Clonal interference and the benefits of sex}
\label{CI}

The reason for the emergence and maintenance of sexual reproduction is a long-standing
puzzle in evolutionary biology, and a number of genetic mechanisms that could explain the ubiquity of sex in higher
organisms have been proposed over the past century. A classic example is the Muller-Fisher mechanism
\cite{Fisher,Muller1932}, which is based on the observation that beneficial mutations arising in different
individuals in an asexual population compete for fixation and therefore obstruct each other's incorporation into
the population; in contrast, in sexuals two individuals carrying different beneficial mutations can mate, thus combining the 
mutations into a single genome. This phenomenon of \textit{clonal interference} sets in when 
the time scale $t_\mathrm{fix}$ of fixation exceeds the time $t_\mathrm{mut}$ between subsequent beneficial mutations,
see Lecture \ref{Intro}, and it is predicted to dramatically slow down the speed of adaptation in large 
asexual populations. 

Early attempts to quantify the Muller-Fisher mechanism arrived at the conclusion that the speed of adaptation reaches
a finite limit for $N \to \infty$ \cite{Crow1965,Felsenstein1974,MaynardSmith1971}, but recent work has uncovered a more
complex scenario \cite{Park2010}. The standard model used in these studies assumes an unlimited supply of beneficial 
mutations with independent fitness effects (no epistatic interactions) and 
selection coefficients $s$ drawn from a probability density $f(s)$. 

Since beneficial mutations typically constitute a small fraction of all possible mutations, there is little empirical
information on the shape of $f(s)$ \cite{EyreWalker2007}, but theoretical arguments favor an exponential form
\cite{Orr2003}; alternatively, for theoretical convenience it is often assumed that all mutations have the same effect
and $f(s) = \delta(s-s_0)$ \cite{Desai2007}. In the latter case a systematic calculation of the speed of adaptation is possible, based on
the idea that the fitness distribution of the population can be described as a traveling wave of constant shape
moving towards higher fitness \cite{Beerenwinkel2007,Rouzine2003,Rouzine2008,Tsimring1996}.  
A key result is that the speed of adaptation is proportional to the logarithm
of population size, in stark contrast to the behavior for small populations where mutations fix independently and 
the dependence is linear in $N$. 

An approximate treatment appplicable to the case of continuous distributions of selection coefficients has been 
proposed by Gerrish and Lenski \cite{Gerrish1998}. This theory assumes that only the mutation with largest
selection coefficient among those appearing during a typical fixation time survives. As a consequence, the speed of adaptation depends
on the tail shape of $f(s)$ and is proportional to $\ln N$ for the exponential distribution.  

Effects of clonal interference on the speed of adaptation have been observed, at least qualitatively, in evolution experiments
with bacterial populations \cite{deVisser1999,deVisser2005,deVisser2006,Elena2003}. By detecting and analyzing individual beneficial
mutations, such experiments can also be used to determine the parameters of the model, primarily the beneficial mutation rate
and the mean selection coefficient \cite{Perfeito2007}; however these estimates depend strongly on the assumption made regarding the distribution $f(s)$ \cite{Hegreness2006}.

As was noted long ago by Maynard Smith \cite{MaynardSmith1968},
the advantage of recombination due to the Muller-Fisher effect
disappears in infinite populations. In that limit recombination
affects the speed of adaptation only if mutations interact
epistatically. To be precise, recombination aids adaptation if the
effect of an mutation decreases as the number of
mutations increases (\textit{negative epistasis}) \cite{Kondrashov1988} but slows it down
in the opposite case. In the presence of \textit{sign epistasis} (as
introduced in Lecture \ref{Acc}) recombination can be strongly
detrimental, leading to a complete localization of the population at
suboptimal fitness peaks for infinite $N$ \cite{deVisser2009,Park2011} and an
exponential growth of the escape time with $N$ when the population
size is finite \cite{Altland2011}. Thus in general recombination can
be beneficial or deleterious depending on the structure of the fitness
landscape. 

\section*{References}

\end{document}